# Very strong coupling in GaAs based optical microcavities

H. Zhang[1], N. Y. Kim[2], Y. Yamamoto[2,3], N. Na[1]

[1] *Institute of Photonics Technologies, National Tsing-Hua University, Hsinchu 30013, Taiwan*
[2] *Edward L. Ginzton Laboratory, Stanford University, Stanford, California 94305, USA*
[3] *National Institute of Informatics, Hitotsubashi, Chiyoda-ku, Tokyo 101-8430, Japan*

We show that when following a simple cavity design metric, a quantum well exciton-microcavity photon coupling constant can be larger than the exciton binding energy in GaAs based optical microcavities. Such a very strong coupling significantly reduces the relative electron-hole motion and makes the polaritons robust against phonon collisions. The corresponding polariton dissociation and saturation boundaries on the phase diagram are much improved, and our calculations suggest the possibility of constructing a room temperature, high power exciton-polariton laser without resorting to wide bandgap semiconductors.

Coherent properties of exciton-polaritons in semiconductors have recently attracted great interests worldwide. Bose-Einstein condensation [1,2], superfluidity [3,4], and vortices [5,6] have been observed, along with a variety of new theoretical concepts [7,8] being proposed. One distinct feature of a polariton condensate is that its effective mass is about 8 orders of magnitude smaller than that of its atomic counterpart. The critical temperature of phase transition may reach up to a room temperature or higher, which opens up exciting opportunities in constructing new classes of coherent optoelectronic devices for real-world applications. With that being said, the effort on studying room-temperature polariton condensate has so far been limited to wide-bandgap semiconductors such as GaN [9,10], ZnO [11] or organic material [12]. While exciton-polaritons in these materials are stable at room temperature because of the large exciton binding energy and the strong photon-exciton coupling strength, significant inhomogeneous broadening is often an issue that strongly obscures the experimental data and interpretation. On the other hand, GaAs, a material that is arguably the cleanest and the most mature platform for semiconductor photonics researches, is seldom considered a candidate for room-temperature polariton condensate despite the fact that strong coupling can be readily reached near room temperature [13,14,15]. This is due to the fact that a quantum well (QW) exciton features a binding energy $E_B \sim 10$ meV comparable to a thermal energy $k_B T \sim 26$ meV at room temperature, and consequently dissociates into free electron-hole pair with high probability when suffering from phonon collisions.

In this paper, we first re-examine a conventional distributed Bragg reflector (DBR) microcavity (MC) and optimize its coupling to multiple QWs. We find that the maximally possible photon-exciton coupling constant $g$ is obtained by carefully engineering the MC *effective refractive index*, and the mode volume plays no role in the end. By applying this design metric, we further propose the use of a guided mode resonance (GMR) MC to reach the full potential of GaAs based optical MCs. $g$ larger than 10 meV can be easily obtained, which surpasses the largest reported value in the literature [16]. Such a strong photon-exciton coupling enforces the system entering the so-called *very strong coupling regime*, first studied by Khurgin *et al.* [20,21], in which the relative electron-hole motion is significantly modified. Phase diagrams of exciton-polaritons in these optimized GaAs MCs are accordingly constructed, and, surprisingly, we find that a dissociation temperature $T_d$ higher than 300 K as well as a saturation density per QW $n_s$ larger than $10^{12}$ cm$^{-2}$ can be reached. In contrast to the common belief, our results suggest that the wide-bandgap semiconductors are not essential for room-temperature polariton condensate. A wide range of practical photonic devices based on GaAs polariton condensates at room temperature would be realizable, benefitting from the superior material quality.

To being with, a conventional DBR MC is considered as shown in Fig. 1. Two configurations are assumed, i.e., InGaAs QWs embedded in a GaAs cavity or GaAs QWs embedded in an AlAs cavity. To maximize the photon-exciton coupling value $g$, we presume that $N$ QWs per antinode are inserted at all electric field antinodes, so that not only in the cavity region but also in the DBR region there are QW excitons coherently coupled to MC photons. The total coupling constant can be calculated by $g = \sqrt{\sum_i g_i^2}$, where $g_i$ is the coupling constant seen by individual QW. The following parameters are used in our calculations: cavity wavelength $\lambda_0$ is 850 nm; Room-temperature GaAs and AlAs refractive index are 3.64677 and 3.00153; 2D oscillator strength $f_{2D}$ of InGaAs and GaAs QW excitons are $4.8 \times 10^{12}$ cm$^{-2}$ [15] and $7 \times 10^{16}$ m$^{-2}$ [16].

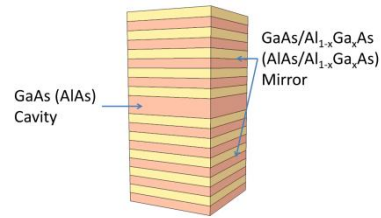

FIG. 1: A conventional DBR MC. QWs are inserted at all electric field antinodes.

We derived analytical approximation (see Supplement

Material) of the total coupling constant for these DBR MCs assuming infinite DBR pairs. Further verifications are performed numerically by transfer matrix method with finite DBR pairs of which its number is set by a condition of peak-to-boundary electric field ratio $E_{peak}/E_{boundary} \sim 10^2$. In Fig. 2, we plot the coupling constant $g$ as a function of external effective length $L_{ext}$, which is tuned by the refractive index contrast between the quarter-wave layers in the DBR. Different internal effective lengths $L_{int}$ are also given, which are controlled by the physical length of the cavity. Note that $L_{ext} + L_{int} = L_{eff}$ is defined, where $L_{eff}$ is the mode effective length. It can be seen in Fig. 2 that the results derived from the analytical approach and the numerical method match very well. For the case of GaAs cavity as shown in Fig. 2 (a), the coupling constant goes up when the effective lengths decrease, which is similar to the standard cavity optimization rule used in cavity quantum electrodynamics researches. On the contrary, as shown in in Fig. 2 (b) for the case of AlAs cavity, the coupling constant goes down when the effective lengths decrease.

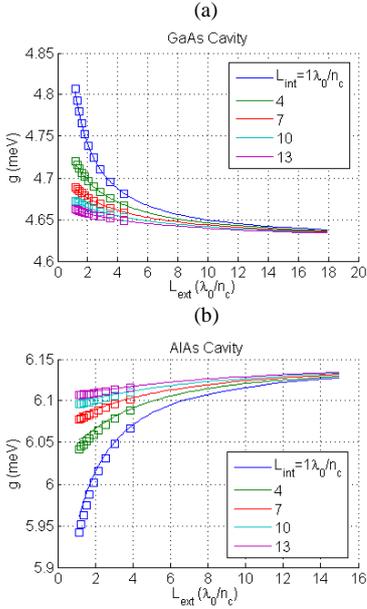

FIG. 2: Coupling constant plotted as a function of external effective length for different internal effective lengths. (a) Cavity region made of GaAs. DBRs are quarter-wave layers made of GaAs and $Al_{1-x}Ga_xAs$. (b) Cavity region made of AlAs. DBRs are quarter-wave layers made of AlAs and $Al_{1-x}Ga_xAs$. Open squares are calculated via numerical method with finite DBR pairs; solid lines are calculated via analytical formula with infinite DBR pairs. Number of QW at each antinode $N=1$ is assumed.

Such a counter-intuitive result can be explained by a simple Fabry-Perot model: assuming a dielectric cavity is sandwiched between two perfect mirrors, $g$ for $N$ QWs inserted per antinode can be derived as

$$g = \sqrt{\frac{Ne^2 f_{2D}}{n_c \varepsilon_0 m_e \lambda_0}}. \quad (1)$$

$n_c$ is the cavity refractive index. Note that $g$ in this Fabry-Perot model is independent of mode effective length because an increase of cavity size ($g$ reduced) implies more QWs can be accommodated ($g$ enhanced) and so the net effect is cancelled. This is true only in a 1D MC and the change of $g$ in Fig. 2 is simply a reflection of adjusting the mode effective index. To more clearly illustrate the concept, we have all coupling constants calculated by transfer matrix method in Fig. 2 re-plotted in Fig. 3. Mode effective indices are then calculated using Eq. (1). A clear correlation can be seen that the saturation at lower/higher $g$ in GaAs/AlAs cavity when $L_{eff}$ increases is determined by the larger/smaller $n_{eff} \approx n_c$.

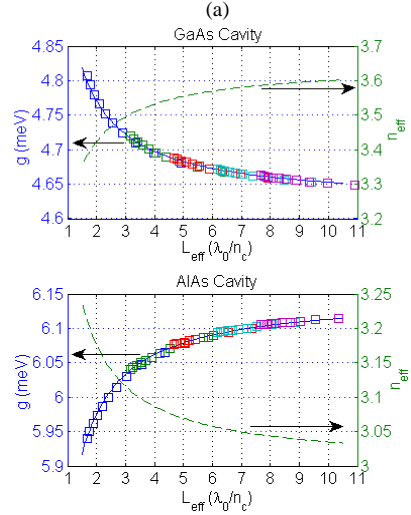

FIG. 3: Coupling constant (open squares) plotted as a function of mode effective length. Blue-solid lines/green-dashed lines are fitting curves to coupling constant/effective index. (a) Cavity region made of GaAs. (b) Cavity region made of AlAs.

With this simple cavity design metric, i.e., a small mode effective index is the key to reach a maximal $g$, we propose using GMR mirrors with air gap in between to form a suspended MC. An exemplary structure is shown in Fig. 4, where GMR is coupled by a 1D photonic crystal through-etch pattern. It is well known that a GMR mirror utilizes Fano-like resonance [17] and can be highly reflective. Such a photonic crystal membrane is a potential candidate to replace the thick DBR layers in a vertical cavity surface emitting laser (VCSEL) [18]. For a proof-of-principle purpose, we use finite-different time domain (FDTD) simulation to design suitable GMR MCs. Given a TE-polarized wave, the period $\Lambda$, space $a$, and thickness $t$ are optimized at 780 nm, 476 nm and 200 nm, respectively. The distance $d$ between the two GMR mirrors is fine tuned to position $\lambda_0$ to 850 nm. Note that all structures in our simulations feature quality factors $Q > 10^4$. In Fig. 5, the electric field $E_z$ is plotted given different cavity lengths $L_c$. It can be seen that the majority of electromagnetic energy is well confined in the air gap region, so a large $g$ is expected according to our studies on conventional DBR MCs. Indeed, the calculated $g$ increases with $L_c$ as the photon "sees" more of the air gap region that is of low refractive index. For example, when $L_c = 3\lambda_0$, $g$ as large as 18 meV given $N=4$ can be obtained, and will ultimately saturate to be 21.3 meV (predicted by Eq. (1) with $n_c=1$) if $L_c$ is further prolonged. Note that in our FDTD simulations presented here, the physical presence of QWs is neglected. We have also optimized the case in which the materials of QW layers are included in the GMR MC. The resultant $g$ and $Q$ change slightly with appropriate design procedures [19], and will not influence the conclusion drawn in this paper.

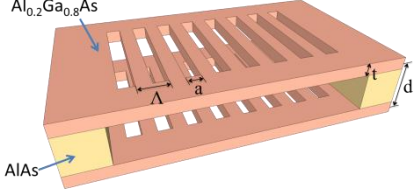

FIG. 4: A proposed GMR MC. An air gap cavity sandwiched by two photonic crystal membrane mirrors is assumed (QWs are not drawn for visual clarity).

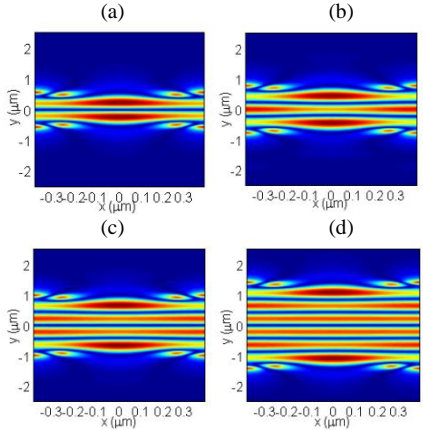

FIG. 5: Distribution of electric field in z direction in one periodic cell of a GMR MC made of $Al_{0.2}Ga_{0.8}As$/air with $L_c$ equal to (a) $\lambda_0$ (b) $1.5\lambda_0$ (c) $2\lambda_0$ (d) $3\lambda_0$. The corresponding coupling constants are 13 meV, 14.6 meV, 15.8 meV, and 18 meV, respectively. Number of QW at each antinode $N=4$ is assumed.

So far we have shown that a very large $g$ can be obtained by following our design metric: small mode effective refractive index. Since $g$ now may exceed $E_B$ in these optimized GaAs MCs, the effect of very strong coupling [20,21] needs to be taken account into consideration so that the electron-hole Coulomb interaction and photon-exciton dipole coupling can be handled on an equal footing. We apply a variational formalism [20] assuming a polariton is described by the linear superposition of an exciton and a photon, i.e., $|\psi_{po}\rangle = \alpha|\psi_{ex}\rangle + \beta|\psi_{pt}\rangle$. $\alpha$ and $\beta$ are the Hopfield coefficients for exciton and photon. A trial wavefunction used for the relative motion of electron-hole pair is $\psi_{ex}(r) = \sqrt{2/\pi}\lambda/a_0 e^{-\lambda/a_0 r}$, where $\lambda$ represents the reduction factor of 2D exciton Bohr radius $a_0$. The system Hamiltonian can be written as

$$H = H_k + H_{e-h} + H_{ex-pt} + H_{pt}, \quad (2)$$

where $H_k = \hbar^2\nabla_r/2\mu + \hbar^2\nabla_R/2M$ is the kinetic energy from the relative and center-of-mass motions of an electron-hole pair; $H_{e-h} = -e^2/4\pi\varepsilon_r\varepsilon_0 r$ corresponds to the Coulomb interaction in an QW exciton; $H_{ex-pt}$ corresponds to the dipole coupling between a QW exciton and a MC photon; $H_{pt}$ is the MC photon energy. Polariton eigen-energies and eigen-states can be determined by minimizing Eq. (2) with $\alpha$ and $\lambda$ as two variational parameters (see Supplement Material). In Fig. 6 (a) and (b), we plot the exciton fraction and the reduction factor of Bohr radius for lower polariton (LP) as a function of $g$ and MC photon-QW exciton energy detuning $\Delta$. When $g/E_B$ is sufficiently large, an increase of photon fraction so that $\alpha^2<0.5$ at $\Delta=0$ emerges, a prominent feature that is different from standard strong coupling. The 2D exciton Bohr radius is significantly reduced at the same time, approaching the values in wide-bandgap semiconductors. In Fig. 6 (c), the upper polariton (UP) and LP eigen-energies (solid lines) at $\Delta=0$ are calculated as a function of $g$, in which a strong asymmetry between UP and LP can be observed. For comparison purpose, a standard Rabi splitting linearly increasing with $g$ (dashed lines) is also presented in Fig. 6 (c). It is shown that the effect of very strong coupling would efficiently stabilize LP by lowering its eigen-energy at a rate that is faster than the case of standard strong coupling. On the contrary, UP becomes unstable as its eigen-energy converges to the zero energy that corresponds to the case of free electron-hole pair.

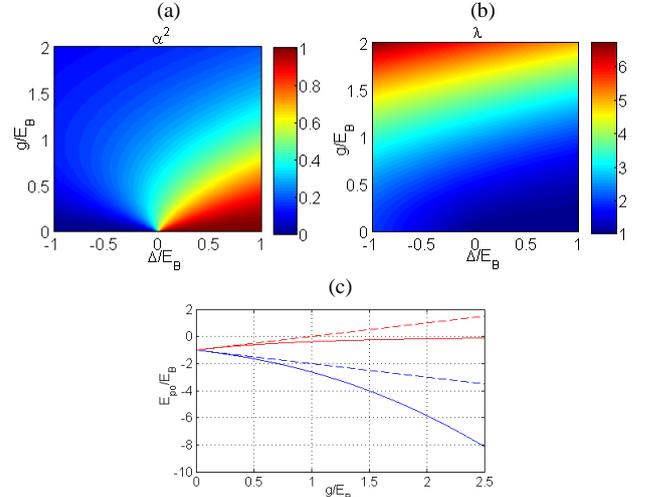

FIG. 6: (a) Exciton fraction of LP plotted as a function of energy detuning and coupling constant. (b) Bohr radius reduction factor of LP plotted as a function of energy detuning and coupling constant. (c) Solid/dashed lines are LP and UP energies calculated with/without variational formalism.

To show our GaAs based optical MCs can be used for room-temperature polariton experiments, we consider the following scenario: within the LP lifetime $\tau_{LP}$, the probability $P$ that phonons may destroy a polariton is proportional to the phonon collision rate $\gamma_{pn}$ associated with the exciton content of a polariton. In addition, an energy at least larger than LP energy $E_{LP}$ minus LP energy half-width $\gamma_{LP}/2$ must be gained by a polariton during its collision with phonons to break into free electron-hole pair. These conditions can be mathematically formulated as

$$P = \alpha^2 \gamma_{pn} \tau_{LP} \left( \frac{1}{2} - \frac{1}{\pi} \arctan \frac{E_{LP}}{\gamma_{LP}/2} \right), \quad (3)$$

assuming the LP lineshape is Lorentzian. For QW excitons in GaAs systems, the phonon collision induced homogeneous broadening can be well modeled by the equation $\gamma_{pn} = \gamma_A T + \gamma_{LO}/(\exp(\hbar\omega_{LO}/k_B T) - 1)$ [15], where $\hbar\omega_{LO}$ (36 meV) is the LO phonon energy. $\gamma_A$ (4.4 μeV) and $\gamma_{LO}$ (15.2 meV) are two phenomenological constants representing the contributions from acoustic and LO phonons. $\tau_{LP} = \hbar/(\alpha^2 \gamma_{ex} + \beta^2 \gamma_{pt})$, where $\gamma_{ex}$ (1.3 μeV) and $\gamma_{pt}$ are the radiative energy full-widths of QW excitons and MC photons. $\gamma_{LP} = \alpha^2 (\gamma_{inh} + \gamma_{pn} + \gamma_{ex}) + \beta^2 \gamma_{pt}$, where $\gamma_{inh}$ (1 meV) is the material inhomogeneous broadening. In Fig. 7 (a), we calculate the disassociation temperature at $\Delta = 0$ by using the condition $P = 0.01$ and assuming a cavity quality factor $Q = 4000$. We find that when $g$ is larger than $E_B = 10$ meV, $T_d$ larger than 300 K becomes possible. The ratio of LP energy to LP energy half-width is also plotted in Fig. 7 (a), and its value coincidently matches well with the literature experimental data [9,10].

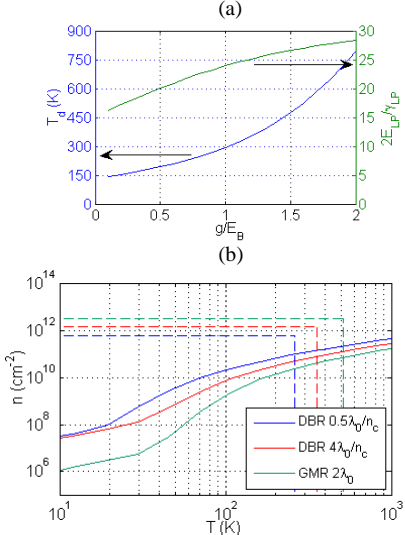

FIG. 7: (a) Dissociation temperature and the ratio of LP energy to LP energy half-width plotted as a function of coupling constant. (b) Polariton phase diagrams for three different structures. Solid curves mark the phase transition from thermal polaritons to polariton BEC. Vertical-dashed lines indicate the polariton thermal dissociation temperatures; horizontal-dashed lines indicate the polariton nonlinear saturation densities. Number of QW at each antinode $N = 4$ is assumed

In Fig. 7 (b), we construct the phase diagrams of exciton-polaritons in GaAs based optical MCs, considering the best experimental structure that exists in the literature [16], an optimized DBR MC, and an optimized GMR MC. The critical density or temperature is calculated by integrating the Bose-Einstein distribution at zero chemical potential along a radially symmetrical in-plane wavevector, with the assumption that the 2D polaritons are trapped in a 50 μm radius laser spot [22]. The vertical dashed lines are the corresponding $T_d$ calculated by setting Eq. (3) to 1%, and the horizontal dashed lines are the corresponding $n_s$ calculated by $n_s/\alpha^2 \approx \lambda^2/(9\pi a_0^2)$ due to phase-space filling and fermionic exchange interaction [23]. For DBR MC with $L_c = 0.5\lambda_0/n_c$ (best structure in the literature), DBR MC with $L_c = 4\lambda_0/n_c$, and GMR MC with $L_c = 2\lambda_0$, the dissociation temperatures are 261K, 355K, and 514K, respectively; the saturation densities per QW are 6.4x10$^{11}$ cm$^{-2}$, 1.4x10$^{12}$ cm$^{-2}$, and 3.4x10$^{12}$ cm$^{-2}$, respectively. $N = 4$ and $\Delta = 0$ are used. The trend is apparent: by following our simple cavity design metric, we may push the system well into the very strong coupling regime, which further increases $\lambda$, lowers $E_{LP}$, and decreases $\alpha$. The net effect significantly improves the saturation and dissociation limits of exciton-polaritons in GaAs based optical MCs, so that an observation of polariton condensate at room temperature becomes accessible. Our results provide a new way to studying macroscopic quantum coherence at room temperature within a clean solid-state environment in which mature fabrication technologies can be applied.

This work is supported by National Science Council (NSC) of Taiwan, Grant No. 101-2112-M-007-002-MY2. Z.H. thanks C.-Y. Hsiao for the preparations of device schematic plots. N.N. thanks National Center for Theoretical Science (NCTS) of Taiwan for partial support.

[1] J. Kasprzak *et al.*, Nature **443**, 409 (2006)
[2] R. Balili *et al.*, Science **316**, 1007 (2007)
[3] A. Amo *et al.*, Nature **457**, 291 (2009)
[4] A. Amo *et al*., Nature Phys. **5**, 805 (2009)
[5] K. G. Lagoudakis *et al.*, Nature Phys. **4**, 706 (2008).
[6] G. Roumpos *et al.*, Nature Phys. **7** 129 (2010)
[7] T. C. H. Liew, A.V. Kavokin, and I. A. Shelykh, Phys. Rev. Lett. **101**, 016402 (2008)
[8] F. P. Laussy, A. V. Kavokin, and I. A. Shelykh, Phys. Rev. Lett. **104**, 106402 (2010)
[9] S. Christopoulos *et al.*, Phys. Rev. Lett. **98**, 126405 (2007)
[10] J. J. Baumberg *et al.*, Phys. Rev. Lett. **101**, 136409 (2008)
[11] M. Zamfirescu *et al.*, Phys. Rev. B **65**, 161205(R) (2002)
[12] S. Kena-Cohen and S. R. Forrest, Nature Photon. **4**, 371 (2010)
[13] S. I. Tsintzos *et al.*, Appl. Phys. Lett. **94**, 071109 (2009)
[14] R. Houdre *et al.*, Phys. Rev. B **49**, 16761 (1994)
[15] A. R. Pratt, T. Takamori, and T. Kamijoh, Phys. Rev. B **58**, 9656 (1998)
[16] J. Bloch *et al.*, Appl. Phys. Lett. **73**, 1694 (1998)
[17] S. S. Wang *et al.*, J. Opt. Soc. Am. A **7**, 1470 (1990)
[18] M. C. Y. Huang, Y. Zhou, and C. J. Chang-Hasnain, Nature Photon. **1**, 119 (2007)
[19] These designs are rather involved and are beyond the scope of this paper. The details will be presented in other publications.
[20] J. B. Khurgin, Solid State Comm. 117 307 (2001)
[21] D. S. Citrin and J. B. Khurgin, Phys. Rev. B **68**, 205325 (2003)
[22] G. Malpuecha *et al.*, Appl. Phys. Lett. **81**, 412 (2002)
[23] S. Schmitt-Rink, D. S. Chemla, and D. A. B. Miller, Phys. Rev. B **32**, 6601 (1985)

# Supplement Material

If the DBR consists of two dielectric material with alternating refractive indices $n_1$ and $n_2$ ($n_1 > n_2$), it can be shown from the dispersion relation of a 1D photonic crystal that the mirror field amplitude attenuates at the rate of $e^{-n_1/n_2 z}$. Assuming the cavity field amplitude is uniform, we may derive an analytical approximation of total coupling constant

$$g = g_0 \sqrt{N} \left( 2s - 1 + \frac{2}{1 - (n_2/n_1)^2} \right)^{0.5}, \text{ for GaAs cavity}$$
$$g = g_0 \sqrt{N} \left( 2s - 2 + \frac{2}{1 - (n_2/n_1)^2} \right)^{0.5}, \text{ for AlAs cavity} \tag{A1}$$

given $N$ QW inserted per antinode. $s$ is the cavity order, and $g_0 = \sqrt{\frac{1}{4\pi\varepsilon_0 n_c^2} \frac{\pi e^2 f_{2D}}{m_e L_{eff}}}$ is the coupling constant of a single QW at the electric field maximum. Effective lengths are defined by

$$L_{eff} = \frac{s - 0.5}{2} \lambda_1 + 2 \frac{n_1^2 f_1 + n_2^2 f_2}{n_1^2 - n_2^2}, \text{ for GaAs cavity i.e. } n_c = n_1$$
$$L_{eff} = \frac{s - 0.5}{2} \lambda_2 + 2 \frac{n_1^2 f_2 + n_1^2 f_1}{n_1^2 - n_2^2}, \text{ for AlAs cavity i.e. } n_c = n_2 \tag{A2}$$

where $f_1 = \frac{-a^2 + 2k_1^2 (\exp(a\lambda_1/4) - 1)}{a(a^2 + 4k_1^2)}$, $f_2 = \frac{a^2 + 2k_2^2 (1 - \exp(-a\lambda_2/4))}{a(a^2 + 4k_2^2)}$, and $a = \frac{8}{\lambda_0} \frac{n_1 n_2}{n_1 + n_2} \ln \frac{n_1}{n_2}$. $\lambda_0$, $\lambda_1$, $\lambda_2$, $k_1$, $k_2$ are free-space wavelength, wavelengths in medium 1 and in medium 2, wavenumbers in medium 1 and in medium 2, respectively.

In the variational formalism, the ground state energy normalized by exciton binding energy can be derived as

$$\frac{E}{E_B} = \alpha^2 \lambda^2 - 2\alpha^2 \lambda - 2\alpha\beta\gamma\lambda + \beta^2 \Delta_0 + \delta(K_R) \tag{A3}$$

where $\gamma = \frac{g}{E_B}$, $\Delta_0 = \frac{\hbar\omega_0 - E_g}{E_B} = -1 + \frac{\Delta}{E_B}$, and $\delta(K_R) = \alpha^2 \frac{\hbar^2 K_R^2}{2ME_B} + \beta^2 \frac{\hbar\omega_0}{E_B} \left( \sqrt{1 + (K_R/K_Z)^2} - 1 \right)$. $g$, $\omega_0$, $E_g$, $\Delta$, $K_R$, and $K_Z$ are coupling constant, cavity frequency, bandgap, photon-exciton energy detuning, in-plane wavenumber, and longitudinal wavenumber, respectively. $E_B = \hbar^2/2\mu a_0^2$ is the exciton binding energy, where $\mu = (m_e^{-1} + m_h^{-1})^{-1}$ and $a_0 = 2\pi\varepsilon_r \varepsilon_0 \hbar^2/\mu e^2$ are the reduced mass and the Bohr radius. $M = m_e + m_h$ is the center of mass. The terms in Eq. (A3) correspond to the kinetic energy of relative electron-hole pair motion, Coulomb interaction, photon-exciton coupling, photon energy, and the kinetic energy of center-of-mass polariton motion. By minimizing the ground state energy, the variation parameters $\lambda$ can be derived as

$$\lambda = 1 + \frac{\beta\gamma}{\alpha} \tag{A4}$$

and the variation parameter $\alpha$ can be derived by

$$(4 + f^2)\alpha^4 - (4 + f^2)\alpha^2 + 1 = 0 \tag{A5}$$

where $f = \left( \gamma^2 - \frac{\Delta}{E_B} + \frac{\hbar^2 K_R^2}{2ME_B} - \frac{\hbar\omega_0}{E_B} \left( \sqrt{1 + (K_R/K_Z)^2} - 1 \right) \right) \Big/ \gamma$. Note that if $f > 0$, $\alpha^2 < 0.5$; if $f \leq 0$, $\alpha^2 \geq 0.5$. In the special case of $K_R = 0$, $\alpha^2 = \frac{1}{2}\left( 1 - \sqrt{1 - 1/(4 + (\gamma^2 - \Delta)^2/\gamma^2)} \right)$ and it's straightforward to show that

$$\frac{E}{E_B} = -\lambda. \tag{A6}$$